\documentclass[10pt,conference]{IEEEtran}
\IEEEoverridecommandlockouts
\makeatletter

\def\endthebibliography{%
\def\@noitemerr{\@latex@warning{Empty `thebibliography' environment}}%
\endlist
}
\makeatother

\usepackage{cite}
\usepackage{amsmath,amssymb,amsfonts}
\usepackage[ruled,vlined]{algorithm2e}    
\usepackage{setspace}
\usepackage{algorithmic}

\usepackage{graphicx}
\usepackage{textcomp}
\usepackage{xcolor}
\def\BibTeX{{\rm B\kern-.05em{\sc i\kern-.025em b}\kern-.08em
	T\kern-.1667em\lower.7ex\hbox{E}\kern-.125emX}}
	
\usepackage[left=0.625in,right=0.625in,top=0.71in,bottom=0.98in]{geometry}
\usepackage{setspace}
\usepackage{multirow}
\usepackage{comment}
\usepackage{tabularx}
\usepackage{booktabs}
\usepackage{graphicx}
\usepackage{subcaption}

\newcolumntype{L}{>{\centering\arraybackslash}m{4.5cm}}
\newcolumntype{K}{>{\centering\arraybackslash}m{2cm}}
\newcolumntype{R}{>{\centering\arraybackslash}m{4.5cm}}

\usepackage{xspace} 
\usepackage[acronym,nogroupskip,nonumberlist,nopostdot]{glossaries}

\newcommand{\vectr}[1]{\boldsymbol{\mathrm{#1}}}
\newcommand{\matr}[1]{\boldsymbol{\mathrm{#1}}}
	
\newcommand\cgaus[2]{\mathcal{CN}(#1,#2)}

\newcommand\norm[1]{\left\lVert#1\right\rVert}

\newcommand{\RomanNumeralCaps}[1]{\MakeUppercase{\romannumeral #1}}

\newcommand{\bG}{\matr{G}}
\newcommand{\bg}{\vectr{g}}
\newcommand{\bY}{\matr{Y}}
\newcommand{\bH}{\matr{H}}
\newcommand{\bQ}{\matr{Q}}

\newcommand{\bA}{\matr{A}}
\newcommand{\bw}{\vectr{w}}
\newcommand{\bq}{\vectr{q}}
\newcommand{\bh}{\vectr{h}}
\newcommand{\bM}{\matr{M}}
\newcommand{\bX}{\matr{X}}
\newcommand{\bx}{\vectr{x}}
\newcommand{\bB}{\matr{B}}

\newcommand{\herm}{\mathsf{H}}

\newcommand{\trp}{\mathsf{T}}

\newcommand{\realset}[2]{ \mathbb{R}^{#1 \times #2}  }
\newcommand{\complexset}[2]{ \mathbb{C}^{#1 \times #2}  }

\newcommand{\re}[1]{\operatorname{Re}\left\{#1\right\} }
\newcommand{\im}[1]{ \operatorname{Im}\left\{#1\right\} }
\newcommand{\tr}[1]{ \operatorname{Tr}\left\{#1\right\} }
\newcommand{\retr}[1]{ \operatorname{ReTr}\left\{#1\right\} }
\newcommand{\rank}[1]{\operatorname{rank} \left( #1 \right) }

\newcommand{\by}{\vectr{y}}

\newacronym{2d}{2D}{two-dimensional}
\newacronym{3d}{3D}{three-dimensional}
\newacronym{4g}{4G}{fourth generation}
\newacronym{5g}{5G}{fifth generation}
\newacronym{5gnr}{5G NR}{5G New Radio}
\newacronym{3gpp}{3GPP}{third-generation partnership project}
\newacronym{adc}{ADC}{analog-to-digital converter}
\newacronym{am}{AM}{amplitude modulation}
\newacronym{ambc}{AmBC}{ambient BC}
\newacronym{ap}{AP}{access point}
\newacronym{ar}{AR}{augmented reality}
\newacronym{aoa}{AOA}{angle-of-arrival}
\newacronym{agc}{AGC}{automatic gain control}
\newacronym{awgn}{AWGN}{additive white Gaussian noise}
\newacronym{bc}{BC}{backscatter communication}
\newacronym{bde}{BD}{backscatter device}
\newacronym{bf}{BF}{beamforming}
\newacronym{ber}{BER}{bit error rate}
\newacronym{bs}{BS}{base station}
\newacronym{bibc}{BiBC}{bistatic BC}
\newacronym{ce}{CE}{carrier emitter}
\newacronym{csi}{CSI}{channel state information}
\newacronym{cvx}{CVX}{convex optimization toolbox}
\newacronym{dac}{DAC}{digital-to-analog converter}
\newacronym{dmimo_s}{D-MIMO}{distributed MIMO}
\newacronym{dl}{DL}{downlink}
\newacronym{dmimo}{D-MIMO}{distributed multiple-input multiple-output}
\newacronym{doa}{DOA}{direction-of-arrival}
\newacronym{dli}{DLI}{direct link interference}
\newacronym{dft}{DFT}{discrete Fourier transform}
\newacronym{dtft}{DTFT}{discrete-time Fourier transform}
\newacronym{en}{EN}{energy neutral}
\newacronym{end}{END}{energy neutral device}
\newacronym{eirp}{EIRP}{effective isotropic radiated power}
\newacronym{etsi}{ETSI}{European Telecommunications Standards Institute}
\newacronym{evd}{EVD}{eigenvalue decomposition}
\newacronym{fdd}{FDD}{frequency-division duplexing}
\newacronym{fdma}{FDMA}{frequency-division multiple access}
\newacronym{fft}{FFT}{fast Fourier transform}
\newacronym{gs}{GS}{grid search}
\newacronym{gd}{GD}{gradient descent}
\newacronym{gsm}{GSM}{Global System for Mobile Communications}  
\newacronym{gna}{GNA}{Girvan-Newman algorithm}
\newacronym{glrt}{GLRT}{generalized log-likelihood ratio test}
\newacronym{iot}{IoT}{Internet-of-Things}
\newacronym{iid}{i.i.d.}{independent and identically distributed}
\newacronym{isr}{ISR}{interference-to-signal ratio}
\newacronym{ieee}{IEEE}{Institute of Electrical and Electronics Engineers}
\newacronym{los}{LoS}{line-of-sight}
\newacronym{lora}{LoRa}{long range}
\newacronym{lti}{LTI}{linear time-invariant}
\newacronym{ls}{LS}{least-squares}
\newacronym{lte}{LTE}{Long-Term Evolution}
\newacronym{lan}{LAN}{local area network}
\newacronym{lsb}{LSB}{least significant bit}
\newacronym{mse}{MSE}{mean square error}
\newacronym{mimo}{MIMO}{multiple-input multiple-output}
\newacronym{mumimo}{MU-MIMO}{multi-user \gls{mimo}}
\newacronym{miso}{MISO}{multiple-input single-output}
\newacronym{mmwave}{mmWave}{millimeter wave}
\newacronym{mmse}{MMSE}{minimum mean square error}
\newacronym{map}{MAP}{maximum a posteriori probability}
\newacronym{mrc}{MRC}{maximum-ratio combining}
\newacronym{mrt}{MRT}{maximum-ratio transmission}
\newacronym{mobc}{MoBC}{monostatic BC}
\newacronym{nr}{NR}{New Radio}
\newacronym{np}{NP}{Neyman-Pearson}
\newacronym{nfc}{NFC}{near-field communication}
\newacronym{nlos}{NLoS}{non-line-of-sight}
\newacronym{ofdm}{OFDM}{orthogonal frequency division multiplexing}
\newacronym{ofdma}{OFDMA}{orthogonal frequency-division multiple access}
\newacronym{ota}{OtA}{over-the-air}
\newacronym{p1}{P1}{Phase \RomanNumeralCaps{1}}
\newacronym{p2}{P2}{Phase \RomanNumeralCaps{2}}
\newacronym{pl}{PL}{path loss}
\newacronym{pana}{PanA}{Panel A}
\newacronym{panb}{PanB}{Panel B}
\newacronym{pgd}{PGD}{projected gradient descent}
\newacronym{ple}{PLE}{path loss exponent}
\newacronym{pcsi}{PCSI}{perfect channel state information}
\newacronym{papr}{PAPR}{peak-to-average power ratio}
\newacronym{pg}{PG}{path gain}
\newacronym{pdf}{PDF}{probability density function}
\newacronym{phy}{PHY}{physical layer}
\newacronym{rcs}{RCS}{radar cross section}
\newacronym{Riss}{RIS}{Reconfigurable intelligent surfaces}
\newacronym{ris}{RIS}{reconfigurable intelligent surface}
\newacronym{riss}{RIS}{reconfigurable intelligent surfaces}
\newacronym{rf}{RF}{radio frequency}
\newacronym{rfid}{RFID}{radio frequency identification}
\newacronym{rms}{RMS}{root mean square}
\newacronym{rss}{RSS}{received signal strength}
\newacronym{rv}{RV}{random variable}
\newacronym{sdma}{SDMA}{space-division multiple access}
\newacronym{sdr}{SDR}{semidefinite relaxation}
\newacronym{si}{SI}{self-interference}
\newacronym{sumimo}{SU-MIMO}{single-user \gls{mimo}}
\newacronym{svd}{SVD}{singular value decomposition}
\newacronym{smc}{SMC}{specular multipath component}
\newacronym{snr}{SNR}{signal-to-noise ratio}
\newacronym{sinr}{SINR}{signal-to-interference-plus-noise ratio}
\newacronym{sir}{SIR}{signal-to-interference ratio}
\newacronym{siso}{SISO}{single-input single-output}
\newacronym{simo}{SIMO}{single-input multiple-output}
\newacronym{tdoa}{TDOA}{time-difference-of-arrival}
\newacronym{toa}{TOA}{time-of-arrival}
\newacronym{tdd}{TDD}{time division multiplexing}
\newacronym{tdma}{TDMA}{time-division multiple access}
\newacronym{ue}{UE}{user equipment}
\newacronym{ul}{UL}{uplink}
\newacronym{uhf}{UHF}{ultra high frequency}
\newacronym{ula}{ULA}{uniform linear array}
\newacronym{upa}{UPA}{uniform planar array}
\newacronym{ura}{URA}{uniform rectangular array}
\newacronym{uwb}{UWB}{ultrawideband}
\newacronym{zf}{ZF}{zero-forcing}
\newacronym{qam}{QAM}{quadrature amplitude modulation}
\newacronym{qos}{QoS}{Quality of Service}
\newacronym{wlan}{WLAN}{wireless local area network}
\newacronym{wpt}{WPT}{wireless power transfer}
\newacronym{wrt}{w.r.t.}{with respect to}

\begin{document}

\title{
	Reducing Dynamic Range in Bistatic Backscatter Communication via Beamforming Design 
	\thanks{This work was funded by the REINDEER project of the European Union's Horizon 2020 research and innovation program under grant agreement No. 101013425, and in part by ELLIIT and the KAW foundation.}}

\author{Ahmet Kaplan, Diana P. M. Osorio, and Erik G. Larsson\\
	\IEEEauthorblockA{Department of Electrical Engineering (lSY), Linköping University, 581 83 Linköping, Sweden.
}}

\maketitle

\begin{abstract}
Considering the exponential growth of Internet-of-Things devices and the goals toward sustainable networks, the complexity should be focused on the infrastructure side. For a massive number of passive devices, \gls{bc} is a promising technology that reduces cost and increases energy efficiency by enabling transmitting information by backscattering radio frequency signals. Two main limitations that restrict the performance of \gls{bc} are the round-trip path loss effect and the \gls{dli} from the \gls{ce}. To circumvent this, we propose a novel transmit beamforming design for a multiple antenna \gls{bibc} system that realizes both purposes: mitigation of the \gls{dli} and increasing the power towards the \gls{bde}.
Additionally, we provide a detector design and the performance is evaluated in terms of the probability of error, for which we also provide a closed-form expression. Finally, simulation results show the superiority of the proposed beamforming design in decreasing \gls{dli} over a benchmark scenario  that considers maximum-ratio transmission.
\end{abstract}

\begin{IEEEkeywords}
	Bistatic backscatter communication, beamforming, direct link interference, internet of things
\end{IEEEkeywords}
\glsresetall

\section{Introduction}

The proliferation of \gls{iot} applications is gradually increasing with the evolution of mobile networks.
These applications impact crucial sectors, including transportation, manufacturing, security, and healthcare.
Therefore, the massive number of constrained \gls{iot} devices urges the development of low-complex, low-cost, and highly energy-efficient solutions. Under this scenario, \gls{bc} 
emerges as a promising technology to provide connectivity and meet these stringent requirements.

Different from traditional communications, \gls{bc} uses external \gls{rf} signals to transmit information, and it can be implemented in three different configurations: \gls{mobc}, \gls{bibc}, and \gls{ambc}. The main elements of a \gls{bc} system are a \gls{ce}, a reader, and a \gls{bde}. In \gls{mobc}, the \gls{ce} and the reader are co-located and can share several parts of the same infrastructure. In \gls{bibc}, the \gls{ce} and reader are spatially separated, which provides flexibility to adjust their locations for optimal performance. 
In addition, \gls{bibc} does not require full-duplex communication, a complex technology that \gls{mobc} requires.  The last type of \gls{bc}, \gls{ambc}, does not have a dedicated \gls{ce} and relies on ambient \gls{rf} signals, such as Wi-Fi signals, to transmit information. As a result, improving the performance of \gls{ambc} using any solution in the \gls{ce} is not possible.

All types of \glspl{bc} suffer from a round-trip path loss effect due to the path losses from the \gls{ce} to the \gls{bde} and from the \gls{bde} to the reader. This causes the received backscattered power at the reader to be too weak compared to the power of the received carrier signal; in other words, there is a strong \gls{dli} from the \gls{ce}. Therefore, the dynamic range of the received signal in the reader can be high. This can reduce the performance of \gls{bc} because the dynamic range of the \glspl{adc} in the reader may not be enough to detect the weak backscattered signal. Besides, the backscattered signal can be shifted to the last bits of the \gls{adc} due to the low \gls{sir} and quantization errors \cite{biswas2021direct}. 
For this reason, the reader would require high-resolution \glspl{adc} to detect the weak backscattered signals, and this leads to a high-power consumption, particularly in multiple-antenna setups. Therefore, the optimal solution is to mitigate the \gls{dli} before the \glspl{adc}, which allows the use of low-resolution \glspl{adc} in the reader.

\gls{bibc} is a promising technique to solve this problem because the \gls{ce} and the reader do not have to be co-located and algorithms requiring a dedicated \gls{ce} can be designed and used. In addition, \gls{mimo} technology offers several advantages to improve the \gls{bc} performance. For instance, it is possible to apply transmit beamforming techniques using multiple antennas to increase the coverage area, mitigate the \gls{dli}, and reduce the probability of error in \gls{bc}. 
Herein, we focus on the \gls{bibc} with a multiple-antenna setup to
mitigate the \gls{dli} and increase the power towards the \gls{bde}.

\gls{dli} cancellation techniques have been investigated for the different configurations of \gls{bc}. Particularly, in \gls{mobc}, \gls{dli} is called  \gls{si}, and \cite{villame2010carrier, brauner2009novel, bharadia2015backfi, hakimi2022sum} propose algorithms and front-end designs to cancel the \gls{si}. However, the \gls{si} cancellation methods for \gls{mobc} are usually complex and need high power consumption.
In \gls{ambc}, the \gls{dli} cancellation has been attained by receive beamforming \cite{duan2019hybrid, guo2018exploiting}, analog \cite{parks2014turbocharging} and digital \cite{guo2022direct} signal processing, frequency shifting in a \gls{bde} \cite{iyer2016inter, zhang2016enabling}, and using null subcarriers and cyclic prefixes in orthogonal frequency division multiplexing systems \cite{elmossallamy2019noncoherent, yang2017modulation}. 
In \gls{bibc}, the \gls{dli} is suppressed by frequency shifting \cite{varshney2017lorea, li2019capacity} and by coding in a \gls{bde} \cite{tao2021novel}.
However, some of the proposed methods cancel the \gls{dli} after the \gls{adc} \cite{guo2022direct,guo2018exploiting, yang2017modulation, tao2021novel} which requires high-resolution \glspl{adc}, and the frequency shifting technique increases the \gls{bde} complexity. In addition, the aforementioned works for \gls{bibc} have not considered the exploitation of \gls{mimo} technology to mitigate the effect of \gls{dli}.

In \cite{kaplan2023direct}, we proposed a transmit beamforming method to cancel the \gls{dli}. However, we assumed that there is no prior information about the presence of a \gls{bde}, thus the main goal was the \gls{dli} cancellation. In this current paper, we go beyond and propose a novel beamforming design that focuses power to the \gls{bde} by limiting the \gls{dli}.
Our contributions are summarized as follows:
\begin{itemize}
\item We propose a \gls{map} detector to estimate the \gls{bde} information bits and derive a closed-form expression for the probability of error for the proposed detector.

\item We propose a novel beamforming design to mitigate the \gls{dli} and increase the receive power at the \gls{bde}.

\item We show that the proposed method has the capability of canceling \gls{dli} completely.

\item We demonstrate that compared to the benchmark scenario, i.e., \gls{mrt}, our proposed method can achieve similar error probability performance in \gls{bc} while reducing \gls{dli}.

\end{itemize}

\textbf{Notation:} 
This paper uses the following notation: $(\cdot)^\trp, (\cdot)^\herm, \re{\cdot},$ and $\im{\cdot}$ denote transpose, Hermitian transpose, real and imaginary parts of a complex signal, respectively. $\tr{\cdot}$ denotes the trace operator. Boldface capital and lowercase letters represent matrices and vectors, respectively, while italic letters represent scalars. $\norm{\bX}$ denotes the Frobenius norm of the matrix $\bX$, and $\norm{\bx}$ denotes the Euclidean norm of the vector $\bx$. $[\bx]_{i:j}$ represents elements from the $i$-th to the $j$-th position of the vector $\bx$. The complex and real fields are denoted by $\mathbb{C}$ and $\mathbb{R}$, respectively.

\section{System Model}

\begin{figure}[tbp]
	\centering
	\includegraphics[width = 0.8\linewidth]{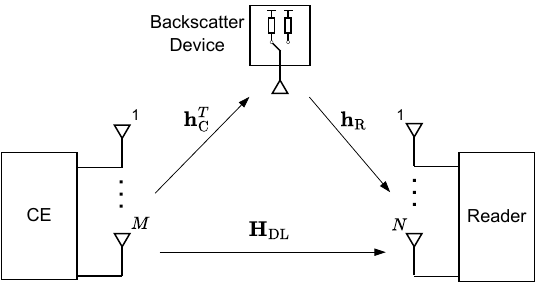}
	\caption{System model of multi-antenna bistatic backscatter communication.}
	\label{fig:System_Model}
\end{figure}
Fig. \ref{fig:System_Model} illustrates a multi-antenna \gls{bibc} system composed by a \gls{ce} with $M$ antennas, a reader with $N$ antennas, and a \gls{bde} with a single antenna. We assume that the \gls{ce} and reader are part of a larger distributed \gls{mimo} setup and have capabilities of a regular \gls{ap} in addition to \gls{bc}.

The \gls{bde} sends its information by changing its reflection coefficient $\gamma$ between two states. We aim to reduce the \gls{dli} and increase the received backscattered power in the reader to detect the \gls{bde} information bits, i.e., the reflection coefficient sequence. For this purpose, we propose a transmit beamforming algorithm maximizing the received backscattered power under a \gls{sir} constraint. 

In Fig.~\ref{fig:System_Model}, $\bH_\text{DL} \in \complexset{N}{M}, \bh_\text{C}^\trp \in \complexset{1}{M},$ and $\bh_\text{R} \in \complexset{N}{1}$ are the channels from \gls{ce} to reader, from \gls{ce} to \gls{bde}, and from \gls{bde} to reader, respectively. We define the backscatter cascade channel as $\bH_\text{BD} = \bh_\text{R} \bh_\text{C}^\trp$, which is a rank-$1$ matrix.

\section{Detector and Performance Analysis}
In this section, we propose and derive the optimal \gls{map} detector for the detection of the \gls{bde} information bits.  The problem is formulated as a hypothesis testing problem where the two scenarios, conditioned on the bit $``0"$ and bit  $``1"$, are considered as hypotheses $\mathcal{H}_{0}$ and $\mathcal{H}_{1}$, respectively as
\begin{equation} \label{eq:hypothesisTesting}
	\begin{split}
		\mathcal{H}_{0}&:  \by_j =\bH_\text{DL} \bx +\gamma_j^0 \bh_\text{R} \bh_\text{C}^\trp  \bx+ \bw_j\\
		\mathcal{H}_{1}&: \by_j =\bH_\text{DL} \bx +\gamma_j^1 \bh_\text{R} \bh_\text{C}^\trp  \bx+ \bw_j,
	\end{split}
\end{equation}
where $\by_j$ is the received signal at the reader in slot $j \in\{1,\dotsc, J\}$ with $J$ denoting the total number of slots. The beamforming vector $\bx$ denotes the transmitted signal by the \gls{ce},
and $\gamma_j^i\ (i=0,1)$ is a real-valued reflection coefficient. 
The vector $\bw_j \in \complexset{N}{1}$ denotes the additive Gaussian noise and all its element are independent and identically distributed $\cgaus{0}{1}$.

We assume that the \gls{ce} and the reader have perfect channel state information, i.e., $\bH_\text{DL}$ and the cascade channel $\bH_\text{BD} = \bh_\text{R} \bh_\text{C}^\trp$ are known. Note that, it is not required to know $\bh_\text{R}$ and $\bh_\text{C}$ separately.

The optimal \gls{map} detector can be expressed as
\begin{equation} \label{eq:map}
	\widehat{\mathcal{H}}_{i}=\underset{\mathcal{H}_{i}}{\operatorname{argmax}}\ P(\mathcal{H}_{i} \mid \matr{Y}) =\underset{\mathcal{H}_{i}}{\operatorname{argmax}}\ p(\mathbf{Y} \mid \mathcal{H}_{i}) P(\mathcal{H}_{i}),
\end{equation}
where $\bY$ is a set of received signals $\by_1, \by_2, \dotsc, \by_J$, and $P\left(\mathcal{H}_{0}\right)$ and $P\left(\mathcal{H}_{1}\right)$ are the prior probabilities. We assume that $P(\mathcal{H}_{0})=P(\mathcal{H}_{1})=1/2$. An equivalent form of the detector given in Eq.~\eqref{eq:map} is given by
\begin{equation}\label{eq:lr}
	L = \frac{ 
		\prod_{j} p\left(\by_j \mid \mathcal{H}_{1}\right)}
	{\prod_{j} p\left(\by_j \mid \mathcal{H}_{0}\right)} \underset{\mathcal{H}_{0}}{\overset{\mathcal{H}_{1}}{\gtrless}} 1,
\end{equation}
where $p\left(\by_j \mid \mathcal{H}_{1}\right)$ and 
$p\left(\by_j \mid \mathcal{H}_{0}\right)$ are \glspl{pdf} of the received signal under $\mathcal{H}_{1}$ and $\mathcal{H}_{0}$, respectively, and given as follows
\begin{equation} \label{eq:pdf}
p\left(\by_j \mid \mathcal{H}_{i}\right) = \frac{1}{\pi^{N}} \exp \left[-\norm{\by_j-\bH_\text{DL} \bx - \gamma_j^i \bH_\text{BD} \bx}^2\right].
\end{equation}

After defining $\by_j^\prime = \by_j-\bH_\text{DL}\bx$, we substitute the \glspl{pdf} into Eq.~\eqref{eq:lr}. Taking the logarithm of both sides, the detector can be expressed as
\begin{equation} 	
	\sum_{j} \left\{\norm{\by_j^\prime - \gamma_j^0 \bH_\text{BD} \bx}^2 - \norm{\by_j^\prime - \gamma_j^1 \bH_\text{BD} \bx}^2\right\} \underset{\mathcal{H}_{0}}{\overset{\mathcal{H}_{1}}{\gtrless}} 0.
\end{equation}
Using the identity
\begin{equation}
		\norm{\bA - \bB}^2 = \norm{\bA}^2 + \norm{\bB}^2 - 2\retr{{\bA}^\herm \bB},
\end{equation}
we can re-express the detector as follows
\begin{equation}\label{eq:final_detector}
\begin{split}
L^\prime &= \sum_{j} (\gamma_j^1 - \gamma_j^0) \re{{\by_j^\prime}^\herm\bH_\text{BD} \bx} \\  &\underset{\mathcal{H}_{0}}{\overset{\mathcal{H}_{1}}{\gtrless}} \sum_{j} \frac{(\gamma_j^1)^2 - (\gamma_j^0)^2}{2}\norm{\bH_\text{BD} \bx}^2 = \mu.
\end{split}
\end{equation}
One can show that $L^\prime$ under $\mathcal{H}_{i}$ has the following distribution
\begin{equation}
		\mathcal{N}\left(\sum\limits_{j} (\gamma_j^1 - \gamma_j^0) \gamma_j^i \norm{\bH_\text{BD} \bx}^2, \sum\limits_{j} \frac{(\gamma_j^1 - \gamma_j^0)^2}{2} \norm{\bH_\text{BD} \bx}^2\right).
\end{equation}

The probability of error is 
\begin{equation} \label{eq:P_e}
	\begin{aligned}
		P_e &= P(\mathcal{H}_{0}) P(\mathcal{H}_{1} | \mathcal{H}_{0}) + P(\mathcal{H}_{1}) P(\mathcal{H}_{0} | \mathcal{H}_{1}) \\
		&= 0.5 (P(L^\prime > \mu | \mathcal{H}_{0}) + P(L^\prime < \mu  | \mathcal{H}_{1}))   \\
		&= Q\left(\frac{\norm{\bH_\text{BD} \bx}}{\sqrt{2}} \sqrt{\sum_j (\gamma_j^1-\gamma_j^0)^2} \right),
	\end{aligned}
\end{equation}
where $Q(x)=\frac{1}{\sqrt{2 \pi}} \int_x^{\infty} \exp \left(-\frac{u^2}{2}\right) d u$. As seen in Eq.~\eqref{eq:P_e}, $P_e$ decreases with increasing $\norm{\bH_\text{BD} \bx}$. Therefore, in the next section, we define our problem to maximize $\norm{\bH_\text{BD} \bx}$ in order to minimize $P_e$ under a constraint on \gls{sir} which is necessary due to the hardware limitations.  \footnote{Note that Eq. \eqref{eq:P_e} is derived assuming infinite resolution \glspl{adc}. In practice, quantization errors will increase $P_e$.}

\section{Proposed Beamforming Method}

The dynamic range of an $n$-bit \gls{adc}, given by $6.02n$ dB \cite{baker2008cmos}, must be proportional to the signal strength ratio between the \gls{dli} and the received weak backscattered signal. Therefore, when optimizing error performance by designing the transmit beamforming vector, we should also consider the \gls{sir} due to the hardware limitations on the \glspl{adc} dynamic range in the reader circuitry. This is because the dynamic range of the received signal, and consequently the quantization error, increases with the decreasing \gls{sir}, which is the ratio of the received powers from the backscattered signal and the direct link expressed as
\begin{equation}
	\text{SIR} = \frac{1}{\eta} = \frac{\norm{\bH_\text{BD} \bx}^2}{\norm{\bH_\text{DL} \bx}^2},
\end{equation}
where $\eta$ is the dynamic range of the received signal.

Therefore, we can formulate our optimization problem as
\begin{equation} \label{eq:maximization}
	\begin{aligned}
		\mathcal{O} \mathcal{P}_{\text{C}}: \quad & \underset{\bx\in \complexset{M}{1}}{\text{max}}
		& & \norm{\bH_\text{BD}  \bx}^2 \\
		& \text{subject to}
		& & \frac{\norm{\bH_\text{DL} \bx}^2}{\norm{\bH_\text{BD} \bx}^2} = \eta \leq \alpha, \norm{\bx}^2 \leq P_{max}, \\
	\end{aligned}
\end{equation}
where $\alpha$ is the constraint on $\eta$, and the required dynamic range of the \glspl{adc} in the reader is proportional to $\alpha$. The parameter $\alpha$ can be chosen based on the requirement of specific use cases and resolution of the \gls{adc} in the reader. As the $\alpha$ value increases, the received backscattered power also increases at the cost of increasing \gls{dli}. For instance, when $\alpha$ is sufficiently large, the proposed beamforming technique corresponds to \gls{mrt}, while complete cancellation of \gls{dli}  occurs for $\alpha = 0$. 

\subsection{Transformation into the Real Domain}
For simplicity, we present the equivalent transformation of the problem described in Eq.~\eqref{eq:maximization} into the real domain. \footnote{Note that the problem can also be solved in the complex domain.} First, we define the following matrices

\begin{equation} 
	\allowdisplaybreaks
	\begin{split}
		\bG_\text{BD} &\triangleq
		\begin{bmatrix}
			\re{\bH_\text{BD}} & -\im{\bH_\text{BD}} \\
			\im{\bH_\text{BD}} & \re{\bH_\text{BD}}
		\end{bmatrix},
		\\
	    \bG_\text{DL} &\triangleq
		\begin{bmatrix}
			\re{\bH_\text{DL}} & -\im{\bH_\text{DL}} \\
			\im{\bH_\text{DL}} & \re{\bH_\text{DL}}
		\end{bmatrix},			
		\bx^\prime \triangleq
		\begin{bmatrix}
			\re{\bx} \\ \im{\bx}
		\end{bmatrix}.
	\end{split}
\end{equation}
The problem $\mathcal{O} \mathcal{P}_{\text{C}}$ can be expressed in the real domain by replacing $\bH_\text{BD}, \bH_\text{DL},$ and $\bx$ with  $\bG_\text{BD}, \bG_\text{DL},$ and $\bx^\prime$, respectively.
Using the equality $\norm{\bA}^2 = \tr{\bA \bA^\trp}$ and the cyclic property of the trace operator, the problem can be expressed as
\begin{equation} \label{eq:max_real_trace}
	\begin{aligned}
		\mathcal{O} \mathcal{P}_{\text{R}}: \quad & \underset{\bx^\prime \in \realset{2M}{1}}{\text{max}}
		& & \tr{\bG_\text{BD}^\trp \bG_\text{BD} \bx^\prime {\bx^\prime}^\trp} \\
		& \text{subject to}
		& & \frac{\tr{\bG_\text{DL}^\trp \bG_\text{DL} \bx^\prime {\bx^\prime}^\trp}}{\tr{\bG_\text{BD}^\trp \bG_\text{BD} \bx^\prime {\bx^\prime}^\trp}} \leq \alpha, \\ 
		& & & \tr{\bx^\prime  {\bx^\prime}^\trp} \leq P_\text{max}. \\
	\end{aligned}
\end{equation}

\subsection{Semidefinite Relaxation and Proposed Solution}
The problem $\mathcal{O} \mathcal{P}_{\text{R}}$ is non-convex. Therefore, we use \gls{sdr} to address it, thus resulting in the following problem
\begin{equation} \label{eq:sdr}
	\begin{aligned}
		\mathcal{O} \mathcal{P}_{\text{S}}: \quad & \underset{\bX^\prime \in \realset{2M}{2M}}{\text{max}}
		& & \tr{\bM_\text{BD} \bX^\prime} \\
		& \text{subject to}
		& & \tr{(\bM_\text{DL}-\alpha \bM_\text{BD}) \bX^\prime} \leq 0, \\
		& & & \tr{\bX^\prime} \leq P_\text{max}, \bX^\prime \succeq 0,
	\end{aligned}
\end{equation}
where $\bM_\text{BD}$$=$$\bG_\text{BD}^\trp \bG_\text{BD}$, $\bM_\text{DL}$$=$$\bG_\text{DL}^\trp \bG_\text{DL}$, and $\bX^\prime$$=$$\bx^\prime {\bx^\prime}^\trp$ are positive semidefinite matrices and $\rank{\bM_\text{BD}}$$=$$1$.

The problem $\mathcal{O} \mathcal{P}_{\text{S}}$ is a convex problem and has a global optimal solution. It can be solved using MATLAB-based \gls{cvx}. The global optimal solution of the problem $\mathcal{O} \mathcal{P}_{\text{S}}$ is denoted as $\bX^\prime_\text{opt}$, and its eigenvalue decomposition is given by
\begin{equation}
	\bX^\prime_\text{opt} = \bQ \mathbf{\Lambda} \bQ^\trp,
\end{equation}
where the columns of $\bQ \in \realset{2M}{2M}$ and the diagonal elements of $\mathbf{\Lambda} \in \realset{2M}{2M}$ consist of the eigenvectors of $\bX^\prime_\text{opt}$ and the eigenvalues of $\bX^\prime_\text{opt}$ arranged in descending order, respectively. 
The best rank-$1$ approximation of $\bX^\prime_\text{opt}$ is given by $\lambda_1 \bq_1 \bq_1^\trp$, where $\lambda_1$ is the first diagonal element of $\mathbf{\Lambda}$, $\bq_1$ is the first column of $\bQ$ and $\norm{\bq_1}$$=$$1$ \cite{eckart1936approximation}.
The solution to the problem $\mathcal{O} \mathcal{P}_{\text{R}}$
is the scaled version of the dominant eigenvector of $\bX^\prime_\text{opt}$ as follows:
$
	\bx^\prime_\text{sol} = \sqrt{P_\text{max}}\bq_1.
$
Finally, the solution for $\mathcal{O} \mathcal{P}_{\text{C}}$ is given as 
\begin{equation}
	\bx_\text{sol} = [\bx^\prime_\text{sol}]_{1:M} + j[\bx^\prime_\text{sol}]_{M+1:2M}.
\end{equation}

\section{Numerical Results}
In this section, we evaluate the proposed beamforming design for some illustrative cases. We use the following parameters for the simulations: $M=16, N=16, \lambda=0.1 \text{ m}, P_\text{max}=1, J=1, \gamma_j^0=-1, \gamma_j^1=1$ and the inter antenna distances are $0.5 \lambda$ both in \gls{ce} and reader. The term $\lambda$ stands for the wavelength of the emitted signal.
The \gls{ce} and reader are centered at the positions $(x,y,z) = (0,0,0)$ and $(0,8,0)$, respectively. Unless otherwise stated, the \gls{bde} is located at position $(0,2,0)$. There are two reflectors along the $y$-$z$ axis and they are located at $x=2$ and $x=-2$ m. 
The amplitude gain of the \glspl{smc} generated by specular reflections is $g_\text{SMC} = 0.5$.

\begin{figure}[tbp]
	\centering
	\includegraphics[width = 0.9\linewidth]{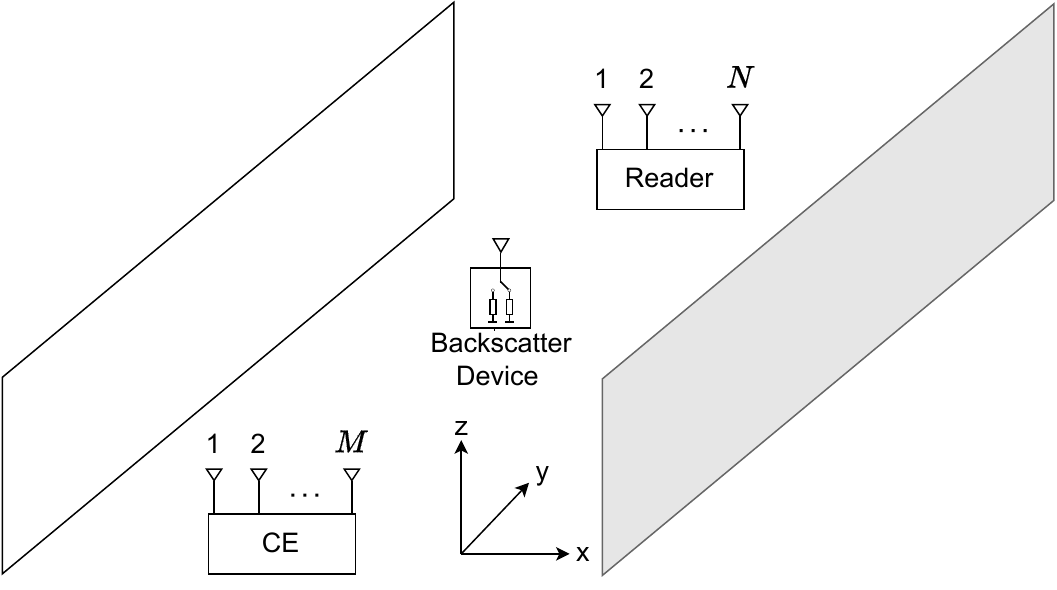}
	\caption{The system model used in the simulations.}
	\label{fig:simulation_model}
\end{figure}

Fig.~\ref{fig:simulation_model} illustrates the deployment of the system for the simulations. As seen in the figure, there are two reflectors at $x$$=$$2$ and $-2$ m. The channels are modeled as \cite{balanis2016antenna, tse2005fundamentals}
\begin{equation}
	\allowdisplaybreaks
	\begin{split}
		[\bH_\text{DL}]_{n,m} &= \frac{\lambda}{4 \pi d_{m,n}} e^{-j \frac{2\pi}{\lambda} d_{m,n}} + \sum_{l=1}^{2} \frac{g_\text{SMC}\lambda}{4 \pi d^l_{m,n}} e^{-j \frac{2\pi}{\lambda} d_{m,n}^l}, \\
		[\bh_\text{C}]_m &= \frac{\lambda}{4 \pi d_{m}} e^{-j \frac{2\pi}{\lambda} d_{m}} + \sum_{l=1}^{2} \frac{g_\text{SMC} \lambda}{4 \pi d^l_{m}} e^{-j \frac{2\pi}{\lambda} d_{m}^l},
	\end{split}
\end{equation}
where $m\in \{1,2,\dotsc,M\}$ and $n\in \{1,2,\dotsc,N\}$.
The distances $d_{m,n}$ and $d_m$ stand for the free-space \gls{los} path lengths between the $m$-th antenna in \gls{ce} and $n$-th antenna in reader, and $m$-th antenna in \gls{ce} and the \gls{bde} antenna, respectively. 
The distances $d^l_{m,n}$ and $d^l_{m}$ stands for the non-\gls{los} path lengths due to the first-order reflections.
The channel $[\bh_\text{R}]_n$ is defined similar to $[\bh_\text{C}]_m$, but using the distances between $n$-th antenna in reader and the \gls{bde} antenna, i.e., $d_{n}$ and $d^l_{n}$. Given that the elements of the noise vector have a unit variance, the \gls{snr} is defined as $\text{SNR}=P_\text{max} J \norm{\bh_\text{R} \bh_\text{C}^\trp}^2 / (MN)$.

\begin{figure}[tbp]
	\centering
	\includegraphics[width = 0.8\linewidth]{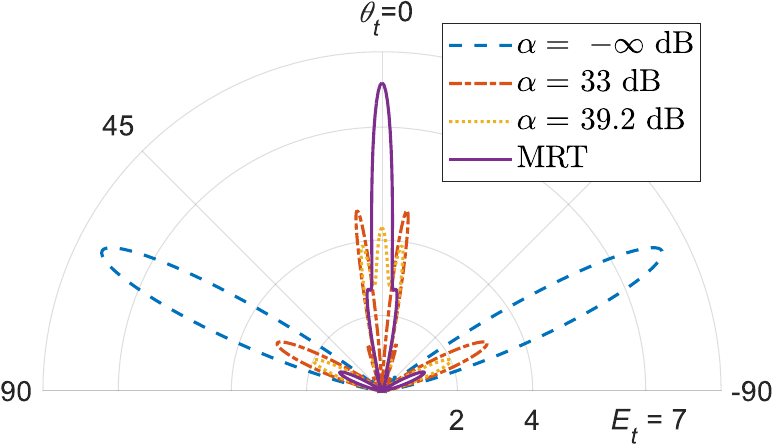}
	\caption{The antenna radiation pattern for the proposed beamforming technique and MRT.}
	\label{fig:radiation_pattern}
\end{figure}

In Fig. \ref{fig:radiation_pattern}, the antenna radiation patterns of \gls{ce} in $x$$-$$y$ axis for $z=0$ are given for \gls{mrt} and the proposed beamforming method with $\alpha=-\infty, 33$ and $39.2\ \text{dB}$. For $\alpha = -\infty$ dB, $\norm{\bH_\text{DL} \bx}^2 = 0$ and the \gls{dli} is completely canceled. The radiation pattern is calculated as 
$
	E_t(\theta) = \norm{\bg(\theta)^\trp \bx}^2,
$
where $\bg(\theta) \in \complexset{M}{1}$ is a steering vector \cite{kaplan2023direct} and $\theta$ is the angle of departure of the transmitted signal.
For \gls{mrt}, the beamforming vector is defined as $\bh_\text{C}^* / \norm{\bh_\text{C}^*}$. As seen in the figure, \gls{mrt} focuses the power to the location of \gls{bde} without considering \gls{dli}, while the proposed beamforming method decreases the \gls{dli} by tuning the parameter $\alpha$. However, decreasing $\alpha$, may also decrease the received backscattered power in the reader. 

\begin{figure}
	\centering
	\begin{subfigure}{.244\textwidth}
		\centering
		\includegraphics[width=1\linewidth]{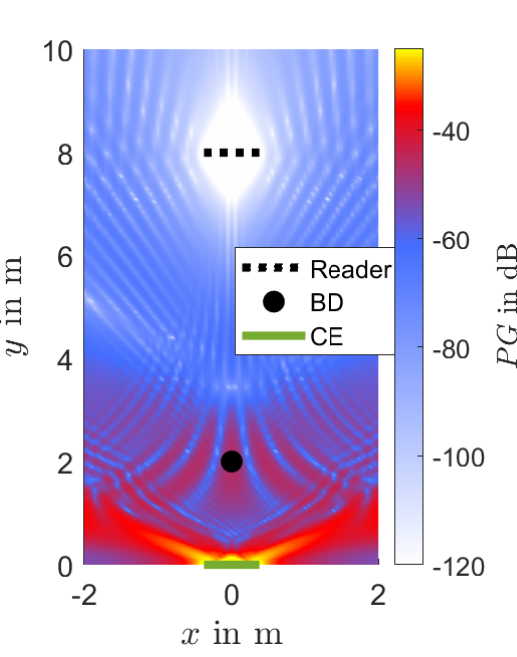}
		\caption{$\alpha = \eta = -\infty$ dB,\\$PG=-45.9$ dB at \gls{bde}}
		\label{fig:sub0}
	\end{subfigure}%
	\begin{subfigure}{.2\textwidth}
		\centering
		\includegraphics[width=1\linewidth]{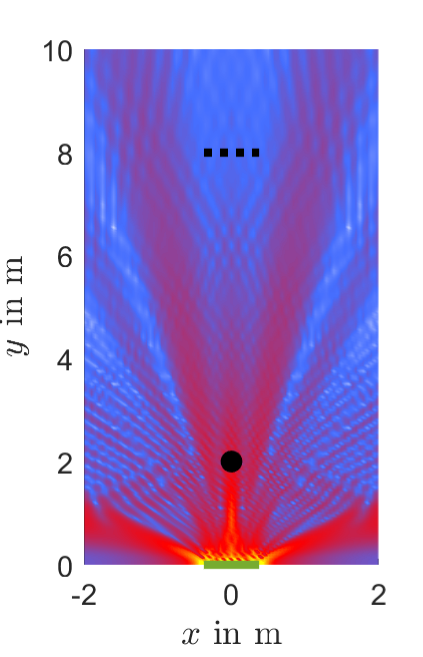}
		\caption{$\alpha = \eta = 33$ dB,\\$PG=-38$ dB at \gls{bde}}
		\label{fig:sub2000}
	\end{subfigure}
		\begin{subfigure}{.202\textwidth}
		\centering
		\includegraphics[width=1\linewidth]{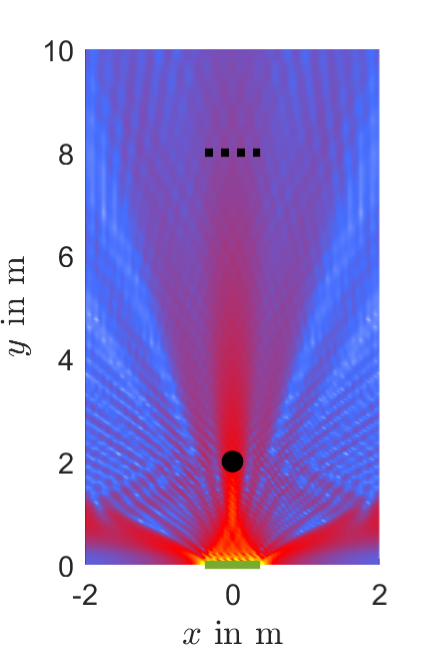}
		\caption{$\alpha = \eta = 39.2$ dB,\\$PG=-35.7$ dB at \gls{bde}}
		\label{fig:sub8000}
	\end{subfigure}
	\begin{subfigure}{.2\textwidth}
		\centering
		\includegraphics[width=1\linewidth]{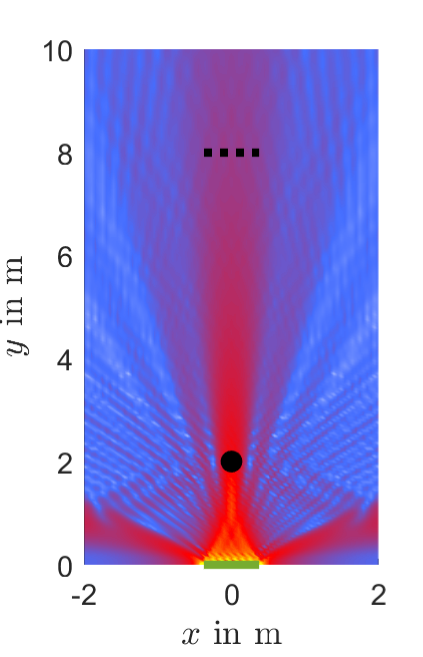}
		\caption{MRT, $\eta=40.9$ dB,\\$PG=-35.5$ dB at \gls{bde}}
		\label{fig:submrt}
	\end{subfigure}
	\caption{Path gains for \gls{mrt} and the proposed beamforming method with different $\alpha$ values.}
	\label{fig:path_gain}
\end{figure}

In Fig.~\eqref{fig:path_gain}, we show the path gain over a rectangular area in $x$$-$$y$ axis when $z=0$. The path gain represents the ratio of the received power to the transmitted power, calculated by
\begin{equation}
	PG = \norm{\bh_\text{C} \bx}^2 /\norm{\bx}^2,
\end{equation}
where $\bh_\text{C}$ shows the channel between transmitter and the location where $PG$ is calculated. As shown in Fig.~\ref{fig:sub0}, for $\alpha=-\infty$ dB, only non-\gls{los} paths are used to transmit power to the \gls{bde}, and the \gls{dli} is completely canceled.
The $PG$s are $-45.9, -38, -35.7$ and $-35.5$ dB at the \gls{bde} location, and $\eta = -\infty,33,39.2,$ and $40.9$ dB for the proposed beamforming method with $\alpha=-\infty, 33, 39.2$ dB and \gls{mrt}, respectively. As shown from these values, both $\eta$ and $PG$ decrease with decreasing $\alpha$ values. In addition, for large enough $\alpha$ values, the proposed beamforming technique will be equivalent to \gls{mrt}, but dynamic range of the received signal will be high. In summary, there is a trade-off between received backscattered power and $\text{SIR}=1/\eta$, and we can control this by adjusting $\alpha$ based on the requirements of the system.

\begin{figure}[tbp]
	\centering
	\includegraphics[width = 0.7\linewidth]{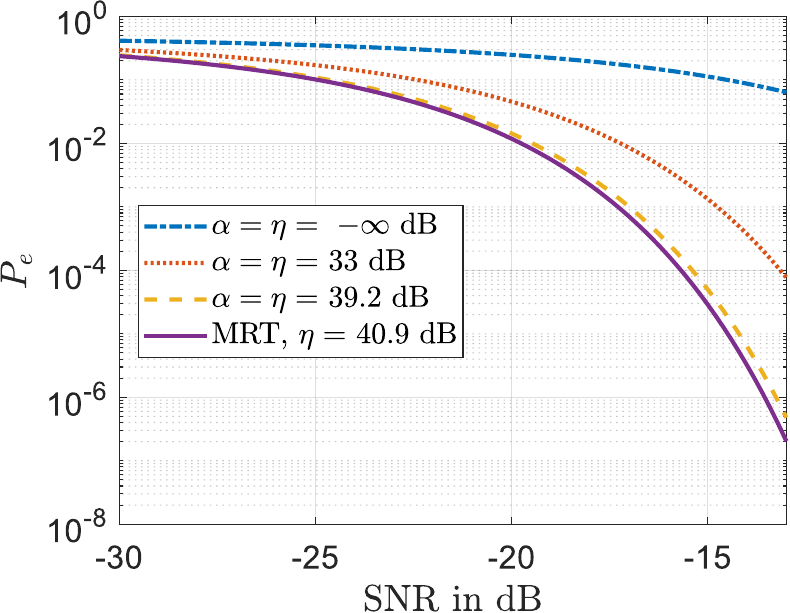}
	\caption{$P_e$ for the proposed beamforming technique and MRT.}
	\label{fig:Pe_BD_Loc_0_2_0}
\end{figure}

\begin{figure}[tbp]
	\centering
	\includegraphics[width = 0.7\linewidth]{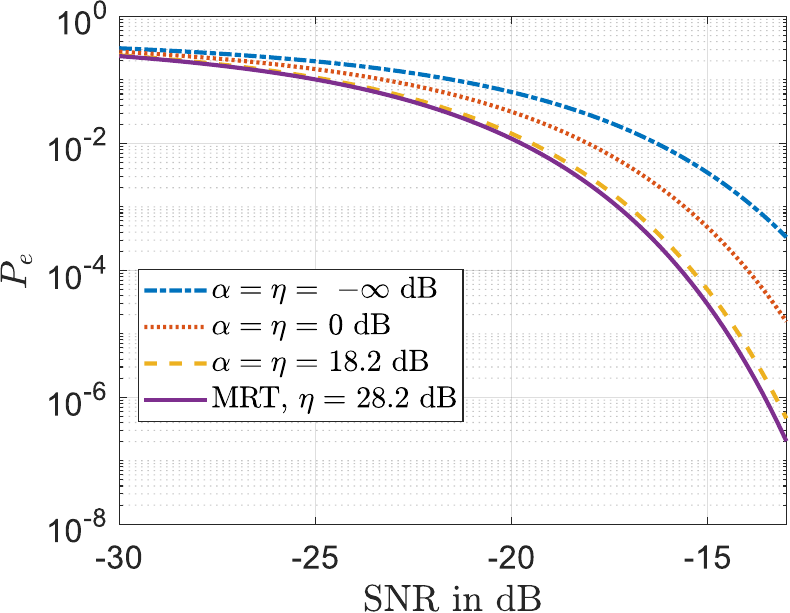}
	\caption{$P_e$ for the proposed beamforming technique and MRT.}
	\label{fig:Pe_BD_Loc_15_2_0}
\end{figure}

In Fig.~\ref{fig:Pe_BD_Loc_0_2_0}, $P_e$ is calculated using Eq.~\eqref{eq:P_e} for \gls{mrt} and the proposed beamforming technique with $\alpha = -\infty, 33,$ and $39.2\ \text{dB}$. With decreasing $\alpha$, while the \gls{dli} decreases, $P_e$ increases because the received backscattered power decreases. For $\alpha=39.2$ dB, while there is no big difference in $P_e$ of the proposed method and \gls{mrt}, there is a $1.7$ dB decrease in $\eta$ $(1.7 \text{ dB increase in SIR})$ by using the proposed method. In the case when we have a limited dynamic range in the reader circuitry, smaller $\alpha$ can be chosen with the cost of increased $P_e$.

In Fig.~\ref{fig:Pe_BD_Loc_15_2_0}, $P_e$ is calculated for \gls{mrt} and the proposed beamforming technique with $\alpha= -\infty, 0$, and $18.2$ dB when \gls{bde} is located at $(1.5,2,0)$ in meters. As shown in the figure, there is a negligible difference in $P_e$ for the \gls{mrt} case and the proposed method with $\alpha=18.2$ dB. While $\eta$ is $28.2$ dB for the \gls{mrt} case,
$\eta$ is $18.2$ dB for the proposed method with $\alpha=18.2$ dB. As a result, there is a $10$ dB decrease in $\eta$ $(10 \text{ dB improvement in SIR})$ by using the proposed beamforming method with negligible difference in $P_e$ compared to the \gls{mrt} case. In addition, to achieve $P_e=10^{-4}$, the required \gls{snr} for the proposed method with $\alpha=18.2$ dB is only $0.3$ dB higher than that of \gls{mrt}, despite a significant $10$ dB reduction in $\eta$.

\section{Conclusion}
This paper studies the \gls{dli} problem in the \gls{bibc} setup with multiple antennas. The detection of the \gls{bde} signal is a challenging problem due to the round-trip path loss effect on the received \gls{bde} signal and the \gls{dli}, which decreases the \gls{sir} and causes quantization errors in the reader circuitry. We propose a novel beamforming design that allows to manage the trade-off between focusing power to the \gls{bde} and cancellation of \gls{dli}. The \gls{sdr} is proposed to solve the optimization problem to find the beamforming coefficients. We also provide an optimal detector for detecting \gls{bde} information bits and derive a closed-form expression for the error probability. The simulation results show that the \gls{dli} is reduced compared to the benchmark scenario, i.e., \gls{mrt}. Therefore, the proposed method allows the use of low-resolution \glspl{adc} on the infrastructure side and has great potential as a solution for energy-efficient massive \gls{iot} networks.

\bibliographystyle{IEEEtran}
\bibliography{references}
\end{document}